\documentclass[12pt,aps,prb,preprint]{revtex4}

\usepackage{amsmath}
\usepackage{graphicx}   

\def\vf{{\varphi}}
\def\be{\begin{equation}}
\def\ee{\end{equation}}
\def\bea{\begin{eqnarray}}
\def\eea{\end{eqnarray}}
\begin{document}

\title{Dimensional Analysis and electric potential due to a uniformly charged sheet}

\author{Amir Aghamohammadi}
 \affiliation{Department of Physics, Alzahra University, Tehran 19384,
IRAN}

 \email{mohamadi@alzahra.ac.ir}
\date{\today}

\begin{abstract}
Dimensional analysis, superposition principle, and continuity of electric potential
are used to study  electric potential of a uniformly charged square sheet at its plane.
It is shown that knowing the electric potential on
the diagonal and inside the square sheet  is equivalent to knowing it everywhere on the plane of square sheet.
Behavior of electric potential near the center of the square is obtained. Then exact solution for the  electric potential at any point on the plane of the square sheet  is obtained. This result is used to calculate the  electric potential of a right triangular sheet on its plane which  can be used to find electric potential at any point on the plane of any uniformly charged polygon sheet.
\end{abstract}

\maketitle

\section{Introduction}
People study physics problems using analytical techniques, approximation methods, and simulations.
Although anybody is interested to have a complete solution for any problem, there are many problems in physics which can not be solved analytically.
To solve such  problems, one may use other methods such as numerical techniques. For the problems which can not be handled analytically, dimensional analysis may also be a useful technique to enlighten the subject, and may give us a better understanding of the subject. This technique despite its simplicity, may give us the opportunity to find some information about the solution, which of course  may be incomplete.
There are some  texts and articles on dimensional analysis, e.g. Refs. (1-6). Dimensional analysis is used in many different areas of physics. In Ref. 6, it is shown that many of the commonly considered applications of weak turbulence possess the incomplete self-similarity property, which can be exploited to obtain core results using a simple dimensional argument.
In ref. 7, dimensional analysis is used to study the resistance force of the fluid that occurs when a body moves through it and the speed of propagation of waves on water.

If one uses dimensional analysis together with some other techniques, it could be more powerful.
One of the standard problems in elementary electricity and magnetism is the electric
field due to an infinite uniformly charged plane. A more physical problem is the electric field of a charged
plane with finite size. Electric potential of the charged sheet at any point can be written as a double integral. Here we are interested in a different method. Our method is based on using dimensional analysis together with superposition principle which comes from linearity of Maxwell equations [8], and symmetries of the system under scaling.
In the first part of this article elementary mathematical techniques is used. Using dimensional analysis, and superposition principle, electric potential due to a uniformly charged square sheet at its plane is studied. In section 2, using dimensional analysis and  superposition principle an identity is obtained between electric potential at the center and the corner of a uniformly charged square sheet. In section 3, some identities are obtained for the electric potential of a uniformly charged square sheet at some different points.
Using these identities one can obtain electric potential everywhere on the plane of square sheet in terms of
the electric potential on the diagonal and inside the square sheet.
In section 4, using the behavior of electric potential at far distances, electric potential at the points near the center of the square sheet is obtained. In section 5, using the behavior of electric potential at large distances, an equation for electric potential is obtained. Solving this equation electric potential on the diagonal of the square sheet is obtained from which electric potential at any arbitrary point on the plane of square sheet is also obtained.
It should be noted that knowing electric potential in the plane of charged sheet is equivalent to knowing electric field in the plane of charged sheet.
In section 6, electric potential at any point on the plane of uniformly charged right triangular sheet is obtained. This helps us to find electric potential at the plane of any uniformly charged sheet polygon.

\section{Dimensional Analysis}
\begin{figure}
\begin{center}
\includegraphics{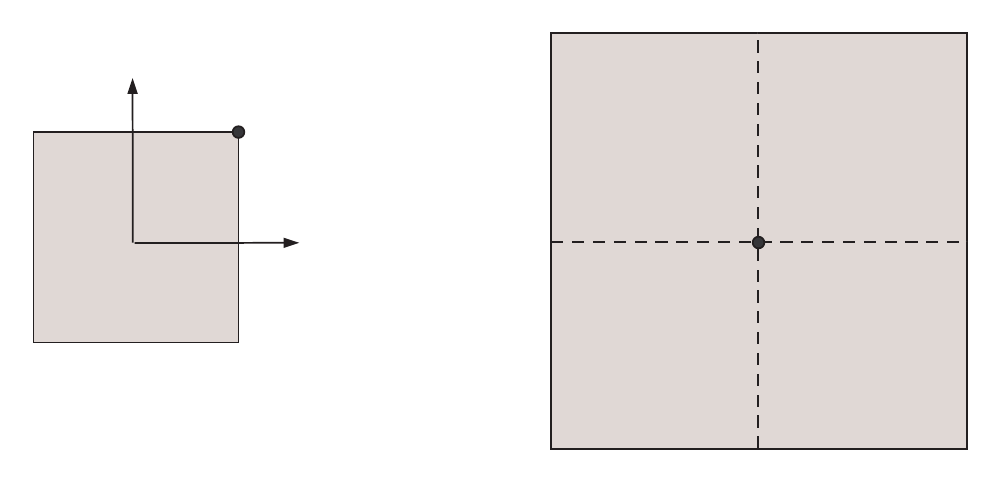}
\setlength{\unitlength}{1mm}
\put(-78,37){$A$} \put(-91,21){$O$}\put(-78,37){$A$} \put(-29,21){$O'$}
\caption{\label{fig:qsquare}}{Using dimensional analysis together with the superposition principle, it can be shown that the electric potential at the center of a square sheet is twice the electric potential at the corner of the uniformly charged square sheet.}
\end{center}
\end{figure}

Let's consider a uniformly charged square sheet with the length $a$.
We are considering the 3D potential, but only on the z = 0 plane, the plane of charged square sheet, and
 cgs units are used.
For a given localized charged distribution, electric potential principally can be obtained uniquely
provided that, electric potential at a reference point is also given. Let's take the potential at infinity
be equal to zero. Then the electric potential of a uniformly charged square at the point $(x,y)$, denoted by
$\Phi(x,y)$, depends on the surface charge density $\sigma$, the length of the square, $a$, and the coordinates $x$, and $y$.
From these parameters, three dimensionless quantity can be constructed
\begin{equation}
\frac{\Phi}{\sigma a},\quad \frac{x}{a},\quad \frac{y}{a}.
\end{equation}
Using dimensional analysis, there is a relation between these dimensionless quantities. The equation between these dimensionless quantities should be such as
\begin{equation}\label{01}
\frac{\Phi(x,y)}{\sigma a}= F(\frac{x}{a},\, \frac{y}{a}),
\end{equation}
where $F$ is an unknown function. Electric potential at the origin $O(x=0, y=0)$, is
\begin{equation}\label{02}
\Phi_O=C_1\sigma a,
\end{equation}
where $C_1= F(0,0)$ is a constant.
It should be noted that $F$ is scale invariant, means that if the lengths  the  of square sheet scales by a factor  $\lambda$, and the coordinates of the point of observations also scales with the same factor, $F$ remains unaltered.
So, if one scales the square sheet with a factor of $2$, remaining $\sigma$ unaltered, electric potential at the center of scaled square sheet becomes twice of the electric potential at the center of unscaled square sheet.
Dimensional analysis cannot tell us anymore.
Electric potential at the point  $\displaystyle{A(x=\frac{a}{2}, y=\frac{a}{2})}$ is
\begin{equation}\label{03}
\Phi(x=\frac{a}{2}, y=\frac{a}{2})=C_2\sigma a,
\end{equation}
where $C_2= F(\frac{1}{2}, \frac{1}{2})$ is another constant.
Because of linearity of Maxwell equations, the electric potential due to a group of charges can be obtained by using
the superposition principle.
 Superposition principle may help us to go further.
Now, let's consider a uniformly charged square with the same charge density $\sigma$, but
whose length is $2a$. Then using (\ref{02}), electric potential at the origin of the
scaled square is
\begin{equation}\label{04}
\Phi'_{O'}= 2a\times \sigma F(0,0)=2C_1\sigma a,
\end{equation}
where $\Phi'_{O'}$ is the electric potential at the point $O'$, the center of the scaled square.
But the point $O'$ is the corner of four squares with the lengths $a$.
See Fig. (\ref{fig:qsquare}).
Using the principle
of superposition,
\begin{equation}\label{004}
\Phi'_{O'}=4 C_2\sigma a.
\end{equation}
Comparing (\ref{04}) and (\ref{004}) one arrives at $C_1=2C_2$, which means that the electric potential at the center of the square is twice the electric potential at the corner of the square.

\section{Some identities on the electric potential of a uniformly charged square sheet}
Consider a uniformly charged square sheet with unit length, and let's choose a point on one of its
diagonals and inside the square sheet, which  divide the diagonal into two parts. Take the ratio of these two parts be  a rational number  $\displaystyle{\frac{n}{m}}$.
We denote the electric potential at this point by $\vf_{n,m}$. Then electric potential at the center of the square is $\vf_{1,1}$, It is obvious that $\vf_{1,1}=\vf_{n,n}$, and $\vf_{n,m}=\vf_{\lambda n,\lambda m}$ for any positive integer $\lambda$. By symmetry it is also obvious that $\vf_{n,m}=\vf_{m,n}$. In this section, we obtain identities between electric potential at different points. Investigating the behavior of
$\lim_{n\to \infty}\vf_{n,n-1}$ gives us electric potential at the points on the diagonal and near the center of the square.

\begin{figure}
\begin{center}
\includegraphics{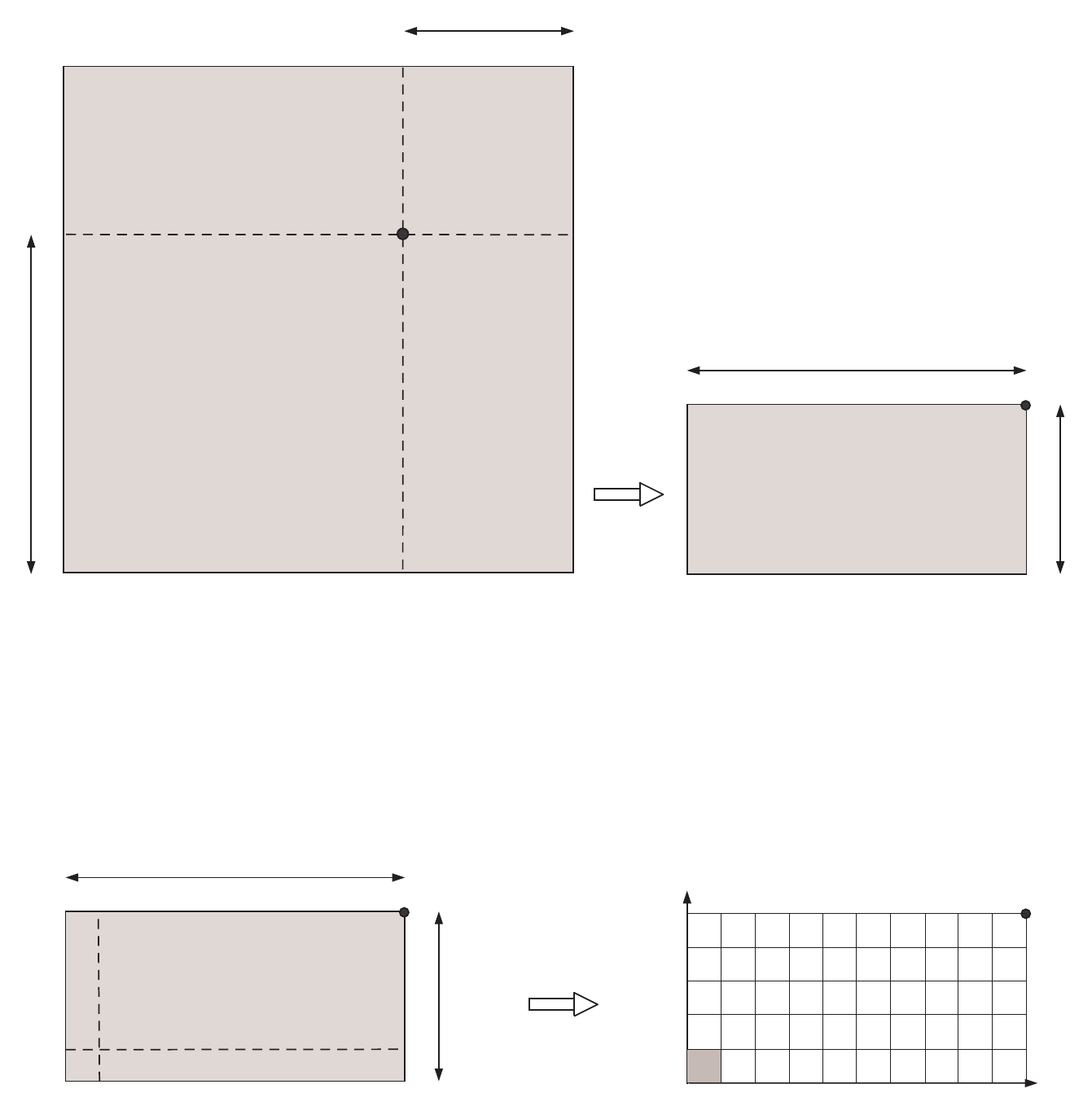}
\setlength{\unitlength}{1mm}
\put(-76,137){$m$} \put(-85,106){$M$}\put(-136,89){$n$}
\put(-31,94){$n$}\put(-3,77){$m$}\put(-6,91){$\psi(n,m)$}\put(-61,110){$(n+m)\vf_{n,m}=\frac{n\ \vf_{1,1}}{2}+\frac{m\ \vf_{1,1}}{2}+2\psi_{n,m}$}
\put(-111,40){$\Phi(n,m)=\left(\psi_{n,m}-\psi_{n-1,m}\right)-\left(\psi_{n,m-1}-\psi_{n-1,m-1}\right)$}
\put(-108,31){$n$}\put(-80,14){$m$}\put(-6,26){$\Phi(n,m)$}
\caption{\label{fig:qsquare2}}Dimensional analysis together with the superposition principle helps us to find a relation between $\Phi(n,m)$ electric potential at any point on the lattice due to square sheet of unit length in terms of four $\psi$ functions, electric potentials of four rectangular sheet at their corners.
\end{center}
\end{figure}

\begin{figure}
\begin{center}
\includegraphics{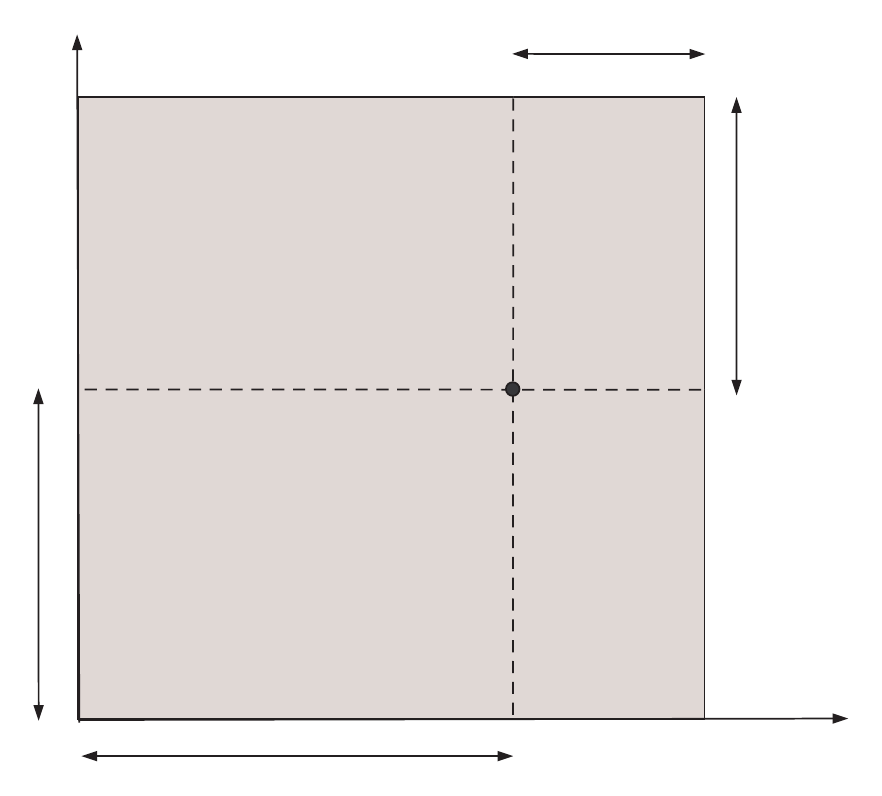}
\setlength{\unitlength}{1mm}
 \put(-61,1){$m$}
\put(-13,56){$k-n$}
\put(-90,25){$n$}
\put(-32,77){$k-m$}\put(-32,77){$k-m$}
\caption{\label{fig:qsquare4}} Electric potential at the any point on a uniformly charged square sheet can be written as a superposition of electric potential of four rectangular sheet.
\end{center}
\end{figure}

Let's consider a uniformly charged square with the length $(n+m)$. As it is $(m+n)$ times of a square with unit length with the same charge density, electric potential at its center is $(n+m)\vf_{1,1}$, and at a point $M$ on its diagonal which  divide it into the ratios of $n$ to $m$ is  $(n+m)\vf_{n,m}$. See Fig.~\ref{fig:qsquare2}.
Using superposition principle, electric potential at the point $M$ can be written as the scalar sum of
\begin{enumerate}
\item The electric potential of a uniformly charged square sheet with the length $n$ at it's corner, which is $\displaystyle{\frac{n}{2}}\ \vf_{1,1}$
\item The electric potential of a uniformly charged square sheet with the length $m$ at it's corner, which is $\displaystyle{\frac{m}{2}}\ \vf_{1,1}$
\item The electric potential of two uniformly charged rectangular  sheet with the lengths $m$, and $n$ at it's corner, where each of them is denoted by $\psi_{n,m}$.
 \end{enumerate}
Then one arrives at
\bea\label{05}
&&(n+m)\vf_{n,m}=\frac{n\ \vf_{1,1}}{2}+\frac{m\ \vf_{1,1}}{2}+2\psi_{n,m},\cr&&\cr
&&\quad \Rightarrow \quad
\psi_{n,m}=\frac{n+m}{2}\left(\vf_{n,m} -\frac{\vf_{1,1}}{2}\right),
\eea
We are now ready to write electric potential due to a uniformly charged square with unit length at any point on a lattice with the same length in the plane of the square sheet. Take the origin at the corner of square.
Let's denote electric potential due to a square sheet with unit length at a point with the coordinates $(n,m)$ (for $n,m>1$), by $\Phi(n,m)$. Then, using superposition principle  $\Phi(n,m)$ is (see Fig.~\ref{fig:qsquare2})
\bea\label{06}
\Phi(n,m)&=&\left(\psi_{n,m}-\psi_{n-1,m}\right)-\left(\psi_{n,m-1}-\psi_{n-1,m-1}\right),\quad \cr
&&\cr
\Phi(n,m)&=&\left( \frac{n+m}{2}\right)\vf_{n,m}-\left( \frac{n+m-1}{2}\right)(\vf_{n-1,m}
+\vf_{n,m-1})\cr &&\cr&&+\left( \frac{n+m-2}{2}\right)\vf_{n-1,m-1}.
\eea
Now, we want to show that electric potential at any arbitrary  point with rational coordinates can be also written in terms of $\vf$'s, electric potential on the diagonal of the square sheet. First consider the points in the square sheet. Let's consider a square with the length $k$, and take $m$, and $n$ two integers less than $k$. See Fig.~\ref{fig:qsquare4}.
It can be divided into four rectangular of the sizes $m\times n$, $n\times (k-m)$, $m\times (k-n)$, and $(k-m)\times (k-n)$.
Electric potential at the point  $(n,m)$ of a uniformly charged square sheet with the length $k$, denoted by $\tilde \Phi(n, m)$, is the sum of electric potential at the corner of these four rectangular
uniformly charged sheets.
\bea\label{08-1}
\tilde \Phi(n, m)&=&\left(\psi_{n,m}+\psi_{n,k-m}+\psi_{k-n,m}+\psi_{k-m,k-n} \right)\cr
&&\cr
&=&\Big( \frac{(k-n+m)}{2}\ \vf_{k-n,m}+\frac{(n+m)}{2}\ \vf_{n,m}+\frac{(k-m+n)}{2}\ \vf_{k-m,n}\cr &&\cr&+&\frac{(2k-n-m)}{2}\ \vf_{k-n,k-m}-k\vf_{1,1}\Big)
\eea
So, electric potential at the point with the coordinate $\displaystyle{(\frac{n}{k}, \frac{m}{k})}$ in a square with the unit length ($m,n <k$) is
\bea\label{09}
\Phi(\frac{n}{k}, \frac{m}{k})&=&\frac{1}{2k}\Big( (k-n+m)\vf_{k-n,m}
+(k-m+n)\vf_{k-m,n} \cr &&\cr &+&(2k-n-m)\vf_{k-n,k-m}+(n+m)\vf_{n,m}-2k\vf_{1,1}\Big)
\eea
This result  can be generalized to any point in the plane of a square sheet of the unit length (inside or outside of it). For this case $m$ or $n$ could be greater than $k$. Then the electric potential at any  point with the coordinate $\displaystyle{(\frac{n}{k}, \frac{m}{k})}$ ($m,n>0$) is
\bea\label{09-0}
\Phi(\frac{n}{k}, \frac{m}{k})&=&\frac{1}{k}\Big[\psi_{n,m}+{\rm sgn}[k-m]\psi_{|k-m|,n}+{\rm sgn}[k-n]\psi_{|k-n|,m}\cr&&\cr
&& +{\rm sgn}[(k-n)(k-m)]\psi_{|k-m|,|k-n|} \Big]
\eea
where ${\rm sgn}[x]$ is the sign function defined through
\be
{\rm sgn}[x]:=\begin{cases}
  +1 & x>0 \\
  -1 & x<0
  \end{cases}.
\ee
Because of the symmetry, electric potential for the points with negative $m$ or $n$ can be obtained easily.
It should be noted that although this solution is obtained for the points
whose coordinates are rational numbers, because of the continuity of the electric potential,
it will be also true for points with real coordinates.
So, if the electric potential of the square sheet at its diagonal is known, electric potential at any other point in the plane of the sheet can be obtained though above mentioned identities.
\section{electric potential near the center of the square}
 Electric potential at the center of a uniformly charged square with unit length, $\Phi(\frac{1}{2},\frac{1}{2})$,  can be calculated analytically.
 \bea
 \Phi(\frac{1}{2},\frac{1}{2})&=&\int_{0}^{1}\int_{0}^{1}\frac{\sigma {\rm d}x{\rm d}y}{\sqrt{(x-1/2)^2+(y-1/2)^2}}\cr &&\cr &=& 4 \sigma \sinh^{-1}(1).
\eea
So  $\vf_{1,1}= 4 \sigma \sinh^{-1}(1)$, and electric potential at the corner of the square sheet is $2\sigma \sinh^{-1}(1)$.
Let's investigate electric potential near the center of square. Electric potential due to a uniformly charged square sheet at the point $(n,n)$ (for far distances, $n\gg 1$) is $\displaystyle{\frac{\sigma}{n\sqrt{2}}}$. So, using equation (\ref{06}), for $n=m$, one arrives at
\bea\label{18}
\Phi(n,n)&=&(2n-1)\vf_{1,1}-(2n-1)\vf_{n-1,n}\cr
&\approx& \frac{\sigma}{n\sqrt{2}}
\eea
Here we have used of $\vf_{n,n}=\vf_{n-1,n-1}=\vf_{1,1}$, and $\vf_{n,n-1}=\vf_{n-1,n}$. The quantity $\vf_{n-1,n}$
for large $n$ is electric potential at a point $A$ near the center of the square and on its diagonal
\bea\label{19}
&&\vf_{n-1,n}\approx \vf_{1,1}-\frac{\sigma}{2\sqrt{2}\ n^2}\cr &&\cr
&&\quad \Rightarrow \quad \Phi(\frac{1}{2}+\epsilon,\frac{1}{2}+ \epsilon)\approx 4 \sigma \sinh^{-1}(1)-4\sigma\ \epsilon^2\sqrt{2}.
\eea
The point $A$ ia at a distance $\epsilon\sqrt{2}$ from the center of square sheet. Electric potential on a point on the plane of square sheet and  near its center can be obtained using  Taylor expansion,
\bea\label{20}
\Phi(\frac{1}{2}+x',\frac{1}{2}+ y')&= &\Phi(\frac{1}{2},\frac{1}{2})+ x'\frac{\partial\Phi}{\partial x'}+
 y'\frac{\partial\Phi}{\partial x'}\cr &&\cr
 &&+ \frac{x'^2}{2}\frac{\partial^2\Phi}{\partial x'^2}+ \frac{y'^2}{2}\frac{\partial^2\Phi}{\partial y'^2}+ x'y'\frac{\partial^2\Phi}{\partial x'\partial y' }+\cdots\cr &&\cr
&= &\Phi(\frac{1}{2},\frac{1}{2})+\alpha x'^2+\beta y'^2+\gamma x'y'+\cdots
\eea
In the above equation, we have used of the fact that electric field in the center of the square sheet is equal to zero. $\alpha$, $\beta$, and $\gamma$ are three constant parameters which is related to second derivatives of electric potential with respect to $x'$, and $y'$. Because of the symmetry of the problem, the points $(\frac{1}{2}+x',\frac{1}{2}+ y')$, and $(\frac{1}{2}+x',\frac{1}{2}- y')$ have the same electric potential, so $\gamma$ should be equal to zero. The points $(\frac{1}{2}+x',\frac{1}{2}+ y')$, and $(\frac{1}{2}+y',\frac{1}{2}+ x')$ have also the same electric potential, so $\alpha=\beta$. So,
\be\label{21}
\Phi(\frac{1}{2}+x',\frac{1}{2}+ y')=\Phi(\frac{1}{2},\frac{1}{2})+\alpha (x'^2+ y'^2)+\cdots.
\ee
 Using (\ref{19}), $\alpha$ should be $-2\sigma\sqrt{2}$.
In summary electric potential near the center of the square has azimuthal symmetry. So, at a point with the distance $\epsilon $ from the center of square sheet, the electric potential is given by
\be\label{22}
\Phi\approx 4 \sigma \sinh^{-1}(1)-2\sigma\ \epsilon^2\sqrt{2}.
\ee
\section{Exact solution for a uniformly charged square sheet}
Here, we want to obtain exact value of electric potential at any point on the plane of square sheet.
It should be noted that knowing electric potential in the plane of charged sheet is equivalent to knowing tangential component of electric field in the plane of charged sheet.  Normal component of the electric field in the charged sheet is $\sigma/2$, and out of the charged sheet it is pure tangential.
The electric potential on the diagonal of square sheet can be obtained. Let's consider equation (\ref{06}) for the point  $n=N\cos \theta$ and $m=N\sin \theta$.
For large N or large distances, the electric potential behaves as $\frac{\sigma}{N}$. In the righthand side of  (\ref{06}), there are four terms. Let's consider each term separately, for large $N$, and up to $\frac{1}{N}$ expansion \bea\label{23}
 \frac{n+m}{2}\ \vf_{n,m}&=&\frac{N}{2}\left(\sin\theta+\cos \theta\right)\vf_{N\cos \theta, N\sin \theta}\cr &=&\frac{N}{2}\left(\sin\theta+\cos \theta\right)\vf_{1, \tan\theta},
\eea
 \bea\label{24}
\frac{n+m-1}{2}\ \vf_{n,m-1}&=&\frac{N(\sin\theta+\cos \theta)-1}{2}\ \vf_{N\cos \theta, N\sin \theta-1}\cr &=&\frac{N(\sin\theta+\cos \theta)-1}{2}\ \vf_{1, \tan\theta-\frac{1}{N\cos \theta}}\cr
&=&\frac{N(\sin\theta+\cos \theta)-1}{2}\left[ \vf_{1, \tan\theta}- \frac{ \vf' }{N\cos \theta} + \frac{\vf''}{2N^2\cos^2 \theta}+\cdots \right],
\eea
\bea\label{25}
\frac{n+m-1}{2}\ \vf_{n-1,m}&=&\frac{N(\sin\theta+\cos \theta)-1}{2}\ \vf_{N\cos \theta-1, N\sin \theta}\cr &=&\frac{N(\sin\theta+\cos \theta)-1}{2}\ \vf_{1, \tan\theta+\frac{\tan\theta }{N\cos \theta}+\frac{\tan\theta }{N^2\cos^2 \theta}+\cdots}\cr&&\cr
&=&\frac{N(\sin\theta+\cos \theta)-1}{2}\left[ \vf_{1, \tan\theta}+ \frac{\tan \theta\ \vf'  }{N\cos \theta} + \frac{ \tan^2\theta\ \vf''}{2N^2\cos^2 \theta}\right. \cr &&+\left.\frac{ \tan\theta\ \vf'}{N^2\cos^2 \theta}+\cdots \right],
\eea
\bea\label{26}
\frac{n+m-2}{2}\ \vf_{n-1,m}&=&\frac{N(\sin\theta+\cos \theta)-2}{2}\ \vf_{N\cos \theta-1, N\sin \theta-1}\cr &=&\frac{N(\sin\theta+\cos \theta)-2}{2}\ \vf_{1, \frac{N\sin\theta-1 }{N\cos \theta-1}}\cr
&=&\frac{N(\sin\theta+\cos \theta)-1}{2}\left[ \vf_{1, \tan\theta}+ \frac{(\tan \theta -1)\ \vf' }{N\cos \theta} + \frac{(\tan\theta -1)^2\vf'' }{2N^2\cos^2 \theta}\right. \cr &&+\left.\frac{(\tan\theta-1)\vf'  }{N^2\cos^2 \theta}+\cdots \right],
\eea
where  prime means differentiation with respect to $\tan \theta$, and $\vf_{1, \tan\theta}$ stands for the potential
at a point on the diagonal of square sheet which divide it with the ratio $1: \tan \theta$.
Gathering all these together up to the term $\frac{1}{N}$, one arrives at
\be\label{27}
  \frac{ \tan \theta }{\cos \theta}\ \vf'_{1,u}+  \frac{\tan\theta (\tan \theta +1) }{2\cos \theta}\ \vf''_{1,u} =-\sigma.
\ee
Defining $u:=\tan \theta$, the above equation recasts to
\be\label{28}
  \frac{u(u+1)}{2}\ \vf''_{1,u}+ u\vf'_{1,u}=-\frac{\sigma}{\sqrt{u^2+1}}
\ee
or
\bea\label{29}
 \frac{{\rm d} }{{\rm d}u}\left(\frac{(u+1)^2}{2}\ \vf'_{1,u}\right)&=&-\frac{\sigma(u+1)}{u\sqrt{u^2+1}}\cr&&\cr
&=&-\frac{\sigma}{\sqrt{u^2+1}}-\frac{\sigma}{u\sqrt{u^2+1}},
\eea
which can be integrated to
\be\label{30}
\vf'_{1,u}=\frac{2\sigma}{(u+1)^2}\{\sinh^{-1}(u)-\sinh^{-1}(\frac{1}{u})+C\},
\ee
where $C$ is constant. Noting that  $\vf'$ is proportional to electric field, and
electric field at the center of square sheet (where $u=1$), vanishes gives $C=0$.
Because of symmetry, electric field on the diagonal and in the plane of square sheet has no component perpendicular to the diagonal
of square sheet. Its component along the diagonal is proportional to $\vf'$, and its component in the $z$ direction
is $\sigma/2$. Then electric field on the diagonal of square sheet is exactly obtained.
Now, electric potential on the diagonal of square sheet can be obtained through an integration,
\be\label{31}
\vf_{1,u}=\int {\rm d}u\ \frac{2\sigma}{(u+1)^2}\{\sinh^{-1}(u)-\sinh^{-1}(\frac{1}{u})\}.
\ee
Using integration by part one arrives at
\bea\label{32}
\vf_{1, u}&=&\frac{2\sigma}{u+1}\{\sinh^{-1}(u)-\sinh^{-1}(\displaystyle{\frac{1}{u})}\}-2\sigma\int \frac{{\rm d}u}{u\sqrt{u^2+1}}\cr&&\cr
&=&2\sigma\left\{ \frac{\sinh^{-1}(u)}{u+1}+\frac{\sinh^{-1}(\displaystyle{\frac{1}{u}})}{1+\displaystyle{\frac{1}{u}}}\right\}+C',
\eea
where $C'$ can be determined using electric potential at the center of square sheet, $\vf_{1,1}=4\sigma\sinh^{-1}(1)$.
Then electric potential  is
\be\label{33}
\vf_{1,u}=2\sigma\left\{ \frac{\sinh^{-1}(u)}{u+1}+\frac{\sinh^{-1}(\displaystyle{\frac{1}{u}})}{1+\displaystyle{\frac{1}{u}}}\right\}
+2\sigma\sinh^{-1}(1).
\ee
As it is seen electric potential at the corner ($u\to \infty$ or $u\to 0$) is $2\sigma\sinh^{-1}(1)$, which is consistent with our previous result.
For the point $(x,x)$, $u= \displaystyle{\frac{x}{1-x}}$, then electric potential at  any point on the diagonal inside the square sheet and with the coordinates  $(x,x)$ is
\be\label{34}
\Phi(x,x)=2\sigma\left\{ (1-x) \sinh^{-1}(\frac{x}{1-x})+x\ \sinh^{-1}(\frac{1-x}{x})\right\}
+2\sigma\sinh^{-1}(1).
\ee
Using (\ref{09}), (\ref{33}), and putting $x:=\frac{n}{k}$, $y:=\frac{m}{k}$, electric potential at any point, $(x,y)$,  in the square sheet, $(0<x,y<1)$, can be written as
\bea\label{35}
\Phi(x,y)&=&\frac{1}{2}\left\{  (1-x+y)\vf_{1-x,y}
+(1+x-y)\vf_{1-y,x}
+(2-x-y)\vf_{1-x,1-y}\right.\cr
&&\left.+(x+y)\vf_{x,y}
-2\vf_{1,1}\right\}\cr
\Phi(x,y)&=&\sigma \left\{ (1-x)\sinh^{-1}(\frac{y}{1-x})+y\sinh^{-1}(\frac{1-x}{y})\right.\cr
&&+ (1-y)\sinh^{-1}(\frac{x}{1-y})+x\sinh^{-1}(\frac{1-y}{x})\cr
&&+ (1-y)\sinh^{-1}(\frac{1-x}{1-y})+(1-x)\sinh^{-1}(\frac{1-y}{1-x})\cr
&&\left.+ x\sinh^{-1}(\frac{y}{x})+y\sinh^{-1}(\frac{x}{y})\right\}.
\eea
Using (\ref{09-0}) electric potential at any arbitrary point, $(x,y)$, ($x,y>0$)  in the plane of square sheet,  is given by
\bea\label{36}
\Phi(x,y)&=&\sigma \left\{ |1-x|\sinh^{-1}(\frac{y}{1-x})+y\sinh^{-1}(\frac{1-x}{y})\right.\cr
&&+ |1-y|\sinh^{-1}(\frac{x}{1-y})+x\sinh^{-1}(\frac{1-y}{x})\cr
&&+ |1-y|\sinh^{-1}(\frac{1-x}{1-y})+|1-x|\sinh^{-1}(\frac{1-y}{1-x})\cr
&&\left.+ x\sinh^{-1}(\frac{y}{x})+y\sinh^{-1}(\frac{x}{y})\right\}
\eea
Because of the symmetry, electric potential for the points with negative $x$ or $y$ can be obtained easily.
\section{Exact solution for a uniformly charged polygon}
\begin{figure}
\begin{center}
\includegraphics{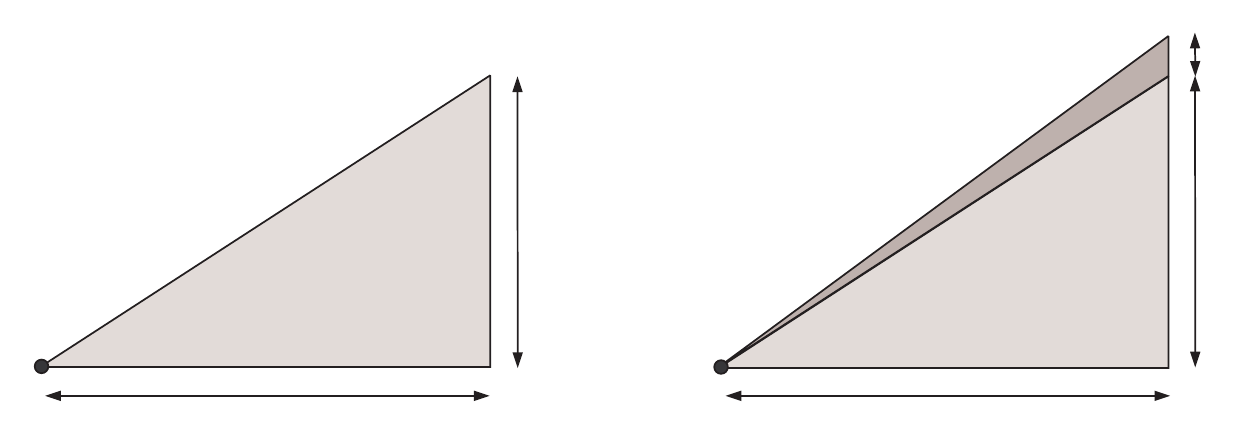}
\setlength{\unitlength}{1mm}
 \put(-99,1){$a$}
\put(-2,38){$\epsilon b$}
\put(-125,5.5){$A$}
\put(-70,20.5){$b$}
 \put(-29,1){$a$}
\put(-56.4,5.5){$A'$}
\put(-2,20.5){$b$}
\caption{\label{fig:qsquare5}} Electric potential at the corner of a uniformly charged right triangular sheet.
\end{center}
\end{figure}

\begin{figure}
\begin{center}
\includegraphics{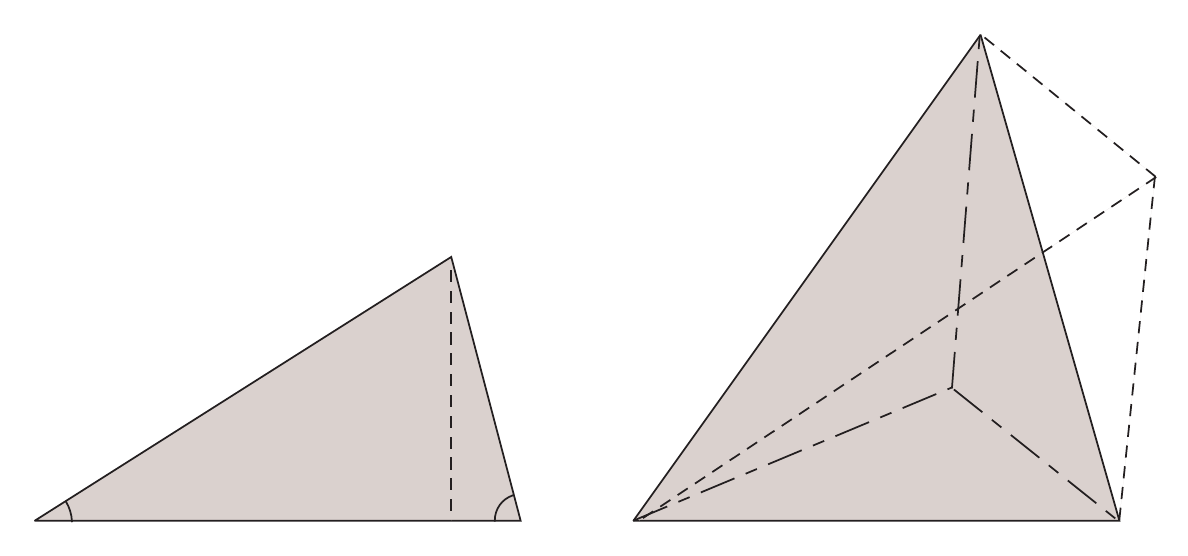}
\setlength{\unitlength}{1mm}\put(-76,33){$A$}
\setlength{\unitlength}{1mm}\put(-90,-1){$a)$}
\setlength{\unitlength}{1mm}\put(-26,12){$B$}
\setlength{\unitlength}{1mm}\put(-30,-1){$b)$}
\setlength{\unitlength}{1mm}\put(-2,37.5){$C$}
\setlength{\unitlength}{1mm}\put(-78,17){$h$}
\setlength{\unitlength}{1mm}\put(-73,6){$\theta_2$}
\setlength{\unitlength}{1mm}\put(-111,4.7){$\theta_1$}
\caption{\label{fig:qsquare6}} $a)$ Electric potential at the corner of a uniformly charged triangular sheet can be written as a superposition of electric potential of two right triangular sheet. $b)$ Electric potential at the point $B$ in the plane  of a uniformly charged triangular sheet can be written as a superposition of electric potential of three triangular sheet, and the electric potential at the point $C$ can be written as the superposition of electric potential of four triangular sheet.
\end{center}
\end{figure}

Using the results for the square sheet, we will obtain some interesting results for any uniformly charged polygon. Let's first consider the a uniformly charged right triangular sheet. It is seen from (\ref{36}) that the electric potential at the corners of a uniformly charged rectangular sheet of the lengths $a$, and $b$ is
\be\label{37}
\sigma [a\,\sinh^{-1}(\frac{b}{a})+b\,\sinh^{-1}(\frac{a}{b})]
\ee
Using (\ref{37}), superposition principle and dimensional analysis the electric potential at the corners of a uniformly charged  right triangular sheet of the length $a$ is
\be\label{38}
{\varPhi}^A_{(a,b)}= \sigma a\sinh^{-1}(\frac{b}{a})+ \sigma a F(\frac{b}{a}),
\ee
where ${\varPhi}^A_{(a,b)}$ is the electric potential at the point $A$ the of a uniformly charged  right triangular sheet of the length $a$ and $b$ (see figure (\ref{fig:qsquare5})), and $F$ is an unknown function.
Using the result for  the electric potential at the corner of a uniformly charged square sheet of the lengths $1$, it can be easily shown that  $F(1)=0$.
Now let's consider a right triangle with lengths $a$, and $b(1+\epsilon)$. The electric potential at the point $A'$ is
the addition of potential due to  triangle with lengths $a$, and $b$, and the electric potential due to a sector.
Electric potential due to a sector of circle with radius $R$, and the angle $\delta \theta$ at the center of circle is
$\sigma R \delta \theta$. So the electric potential at the point $A'$ the corner of a
 a right triangle with lengths $a$, and $b(1+\epsilon)$ up to first order in $\epsilon $ is
\be\label{39}
{\varPhi}^{A'}_{(a,b(1+\epsilon))}\approx \sigma a\sinh^{-1}(\frac{b}{a})+ \sigma a F(\frac{b}{a})+ \epsilon \sigma (\frac{ab}{\sqrt{a^2+b^2}}).
\ee
 Using (\ref{38}) the electric potential at the point $A'$ the corner of the
right triangle  is
\bea\label{40}
{\varPhi}^{A'}_{(a,b(1+\epsilon))}&=& \sigma a\sinh^{-1}\left(\frac{b(1+\epsilon)}{a}\right)+ \sigma a F\left(\frac{b(1+\epsilon)}{a}\right),\cr
&\approx& \sigma a\sinh^{-1}(\frac{b}{a})+ \epsilon \frac{\sigma a b}{\sqrt{a^2+b^2}} +\sigma a F(\frac{b}{a})+\epsilon\sigma b F'(\frac{b}{a}),
\eea
where the $F'$ stands for  differentiation of $F$ with respect to its argument. Comparing (\ref{39}), (\ref{40}) gives $F'=0$ and $F$ is a constant. But $F(1)=0$, so $F$ should  be zero. Then the electric potential at the corner of a right triangular at its corner is $\sigma a\sinh^{-1}(\frac{b}{a})$ (see figure (\ref{fig:qsquare5})).
Electric potential at the vertices of any arbitrary triangle can be obtained easily. See the figure (\ref{fig:qsquare6})).
Electric potential at the point $A$ is the superposition of electric potential of two right triangular sheet.
\bea\label{41}
{\varPhi}^A&=& \sigma h\left[\sinh^{-1}(\cot\theta_1)+\sinh^{-1}(\cot\theta_2)\right]\cr
&=& \sigma h\ln\left(\cot(\frac{\theta_1}{2})\cot(\frac{\theta_2}{2})\right),
\eea
where we have used of $\sinh^{-1}(u)=\ln (u+\sqrt{u^2+1})$. Electric potential at any point in the plane  of a uniformly charged triangular sheet can be written as a superposition of electric potential of three (or four) triangular sheet (if the point is out of triangle) on the plane of the sheet. Similarly electric potential at any point in the plane  of an arbitrary polygon can be written in terms of electric potential of some triangles. This gives us also an upper and a lower bound on the magnitude of electric potential at any point in the plane  of a uniformly charged sheet of any shape. The lower and upper bounds come from the electric potential due to two polygons which are surrounded by  and the one which is surrounded the charged area.

 \begin{acknowledgments}
I would like to thank M. Khorrami, and A. Shariati for useful discussions. I was partially supported by the
research council of the Alzahra University.
\end{acknowledgments}

\newpage


\begin{thebibliography}{5}
\bibitem{PWB} P. W. Bridgman; \textsl{Dimensional Analysis} Yale University, New Haven, 1922.

\bibitem{LIS} L. I. Sedov; \textsl{Dimensional Methods in Mechanics} CRC Press, 1993.

\bibitem{MHH}M. H. Holmes; \textsl{Introduction to the Foundations of Applied Mathematics} Springer, 2009.

\bibitem{CFB} C. F. Bohren; \textsl{Dimensional analysis, falling bodies, and the fine art of not solving
differential equation} Am. J. Phys. {\bf 72} (4), 534--537 (2004).

\bibitem{JFP}  J. F. Pricea; \textsl{Dimensional analysis of models and data sets}
Am. J. Phys. {\bf 71} (5) 437--447 (2003).

\bibitem{CSN} C. Connaughton, S. Nazarenko, \& A. C. Newell; \textsl{Physica D: Nonlinear Phenomena} {\bf 184} (1-4), 86-97 (2003).

\bibitem{MNLN} T. Misic, M. Najdanovic-Lukic \& L. Nesic; \textsl{European Journal of Physics} {\bf 31} (1-4), 893 (2010).

\bibitem{MHN} M. H. Nayfeh, M. K. Brussel; \textsl{Electricity and Magnetism} John Wiley \& Sons, 1985.
\end{thebibliography}
\end{document}